\documentclass[sigconf]{acmart}

\AtBeginDocument{%
  \providecommand\BibTeX{{%
    \normalfont B\kern-0.5em{\scshape i\kern-0.25em b}\kern-0.8em\TeX}}}

\acmConference[SIGIR '21]{ SIGIR '21: ACM SIGIR Conference on Research and Development in Information Retrieval}{July 11--15, 2021}

\usepackage[switch]{lineno}  %
\usepackage{amsmath}
\usepackage{subfig}
\usepackage{amsfonts}
\usepackage{algorithm}
\usepackage{algpseudocode}
\usepackage{mwe}

\usepackage{multirow}
\usepackage{booktabs}
\newtheorem{definition}{Definition}

\usepackage{svg}

\setcopyright{none}

\svgsetup{
	inkscapepath=i/svg-inkscape/
}
\svgpath{{svg/}}

\newcommand\numberthis{\addtocounter{equation}{1}\tag{\theequation}}
\newsavebox\CBox
\def\textBF#1{\sbox\CBox{#1}\resizebox{\wd\CBox}{\ht\CBox}{\textbf{#1}}}

\begin{document}

%%
%% The "title" command has an optional parameter,
%% allowing the author to define a "short title" to be used in page headers.
\title{HGKT : Introducing Hierarchical Exercise Graph for 
	Knowledge Tracing}

\author{Hanshuang Tong}
% \authornote{Most experiments are done during work in AIXUEXI Education Group Ltd.}
\orcid{0000-0002-7443-128X}
\affiliation{%
	\institution{Microsoft Corporation, \\ 
		Beijing, China}
}
\email{hanstong@microsoft.com}

\author{Zhen Wang}
\affiliation{%
	\institution{AIXUEXI Education Group Ltd, Beijing, China}
}
\email{wangzhen@aixuexi.com}

\author{Yun Zhou}
\affiliation{%
	\institution{AIXUEXI Education Group Ltd, Beijing, China}
}
\email{zhouyun@aixuexi.com}

\author{Shiwei Tong}
\affiliation{%
	\institution{School of Computer Science and Technology, University of Science and Technology of China}
}
\orcid{0000-0002-4218-0236}
\email{tongsw@mail.ustc.edu.cn}

\author{Wenyuan Han}
\affiliation{%
	\institution{AIXUEXI Education Group Ltd, Beijing, China}
}
\email{hanwenyuan@aixuexi.com}

\author{Qi Liu}
% \author{J. Doe, Q. Public, J. Schmo\thanks{Corresponding author.}}
\authornote{Corresponding Author}
\affiliation{%
	\institution{School of Computer Science and Technology, University of Science and Technology of China}
}
\email{qiliuql@ustc.edu.cn}

%%
%% The "author" command and its associated commands are used to define
%% the authors and their affiliations.
%% Of note is the shared affiliation of the first two authors, and the
%% "authornote" and "authornotemark" commands
%% used to denote shared contribution to the research.

%%
%% By default, the full list of authors will be used in the page
%% headers. Often, this list is too long, and will overlap
%% other information printed in the page headers. This command allows
%% the author to define a more concise list
%% of authors' names for this purpose.

\iffalse
\renewcommand{\shortauthors}{Trovato and Tobin, et al.}
\fi

%%
%% The abstract is a short summary of the work to be presented in the
%% article.
\begin{abstract}
Knowledge tracing (KT) which aims at predicting learner's knowledge mastery plays an important role in the computer-aided educational system. In recent years, many deep learning models have been applied to tackle the KT task, which have shown promising results. However, limitations still exist. Most existing methods simplify the exercising records as knowledge sequences, which fail to explore rich information existed in exercises. Besides, the existing diagnosis results of knowledge tracing are not convincing enough since they neglect prior relations between exercises. To solve the above problems, we propose a hierarchical graph knowledge tracing model called HGKT to explore the latent hierarchical relations between exercises. Specifically, we introduce the concept of problem schema to construct a hierarchical exercise graph that could model the exercise learning dependencies. Moreover, we employ two attention mechanisms to highlight important historical states of learners. In the testing stage, we present a K\&S diagnosis matrix that could trace the transition of mastery of knowledge and problem schema, which can be more easily applied to different applications. Extensive experiments show the effectiveness and interpretability of our proposed models. 

\end{abstract}
\iffalse
We promise to publish the code used in the paper after acceptance.
\fi
%%
%% The code below is generated by the tool at http://dl.acm.org/ccs.cfm.
%% Please copy and paste the code instead of the example below.
%%
\begin{CCSXML}
	<ccs2012>
	<concept>
	<concept_id>10002951.10003227.10003351</concept_id>
	<concept_desc>Information systems~Data mining</concept_desc>
	<concept_significance>500</concept_significance>
	</concept>
	<concept>
	<concept_id>10003456.10003457.10003527.10003540</concept_id>
	<concept_desc>Social and professional topics~Student assessment</concept_desc>
	<concept_significance>500</concept_significance>
	</concept>
	</ccs2012>
\end{CCSXML}

\ccsdesc[500]{Information systems~Data mining}
\ccsdesc[500]{Social and professional topics~Student assessment}

%%
%% Keywords. The author(s) should pick words that accurately describe
%% the work being presented. Separate the keywords with commas.
\keywords{intelligent education, knowledge tracing, hierarchical graph convolutional network, attention mechanism.}

%% A "teaser" image appears between the author and affiliation
%% information and the body of the document, and typically spans the
%% page.

\iffalse
\begin{teaser}
  \includegraphics[width=\textwidth]{diagram/hgkt_diagram.pdf}
  \caption{Seattle Mariners at Spring Training, 2010.}
  \Description{Enjoying the baseball game from the third-base
  seats. Ichiro Suzuki preparing to bat.}
  \label{fig:teaser}
\end{teaserfigure}
\fi

%%
%% This command processes the author and affiliation and title
%% information and builds the first part of the formatted document.
\maketitle

\iffalse
\begin{figure*}[htb]
	\centering
	\includegraphics[width=\linewidth]{diagram/inspiration.pdf}
	\caption{A toy example of knowledge tracing. A learner reads exercises with different knowledge ($Pythagorean\ Theorem$ and $Number\ Axis\ Properties$) and update his knowledge states. Compared to traditional KT methods (DKT, DKVMN...) which would give same prediction to the exercises (i.e. $e_{7}$,$e_{8}$) with same knowledge ($Coordinate \ Calculation$), our method could predict more precisely since we take the prior support relation between exercises into account. Thus, though $e_{7}$ and $e_{8}$ have the same knowledge $Coordinate\ Calculation$, the prediction results vary since the relations between different exercises are different. Here $s_{1}$ denotes problem schema of $e_{7}$ and $e_{8}$}.
	\label{fig:inspiration}
\end{figure*}
\fi

\begin{figure*}[htb]
	\centering
	\includegraphics[width=\linewidth]{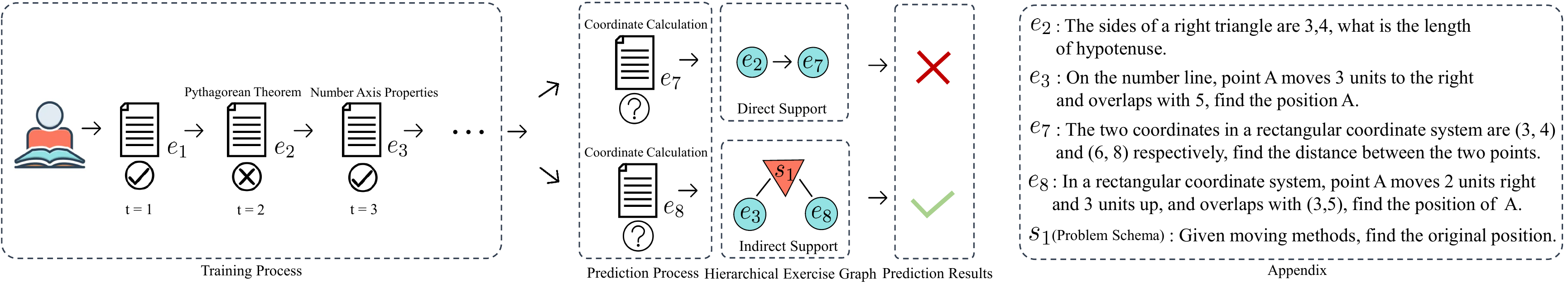}
	\caption{A toy example of knowledge tracing. Given a learner who did exercise $e_{2}$ with knowledge $Pythagorean\ Theorem$ wrong and exercise $e_{3}$ with knowledge $Number\ Axis\ Properties$ right. Suppose we need to predict the learner's performance for two exercises $e_{7}$ and $e_{8}$ which have the same knowledge $Coordinate \ Calculation$. Traditional knowledge tracing method (TKT) such as DKT, GKT would give the same prediction to them since both exercises belong to the same knowledge. However, if we take the prior support relation between exercises (i.e. $e_{2}$ direct support $e_{7}$, $e_{3}$ indirect support $e_{8}$) into account, we can naturally speculate that the learner will more probably do $e_{7}$ wrong and $e_{8}$ right.
	}
	\label{fig:inspiration}
\end{figure*}

\section{Introduction}
Knowledge tracing is a fundamental task in computer-aided educational system, which can benefit both learning and teaching~\cite{anderson2014engaging,burns2014intelligent,jones2004ubiquitous}. A typical knowledge tracing process is as follows: when a question is posted, a learner reads its text and applies knowledge to solve it. After getting a bunch of learners' interactive items, corresponding knowledge sequences and interactive sequences of exercises are extracted to train a KT model which could predict the hidden knowledge states of a learner. However, there are two main problems shown in the traditional knowledge tracing workflow: (1) Exercise representation loss problem: traditional workflow simplifies the exercising records as knowledge sequences, which ignores the difficulty and semantic information contained in exercises. In other words, existing methods have an information loss in the process of representing exercises. (2) Insufficient diagnosis problem: it is also difficult to provide specific learning suggestions based on the diagnosis results. Specifically, suppose we know a learner has a weak knowledge (e.g., $Coordinate\ Calculation$), it is hard to decide which one fits him better (e.g., $e_{7}$ or $e_{8}$ in Fig.~\ref{fig:inspiration}), since mappings between knowledge and exercises is too broad~\cite{hontangas2000choice,zhao2018automatically}.

In the literature, there are many efforts in knowledge tracing. Existing methods can be divided into two tracks: traditional knowledge based and exercise based track. The traditional knowledge based method converts learners'exercising sequences into knowledge sequences without considering the text information of the exercises. The most popular one is Bayesian Knowledge Tracing (BKT)~\cite{BKT}, which updates learner's knowledge state by a Hidden Markov Model. Deep learning methods like Deep Knowledge Tracing (DKT) model the learning process as a recurrent neural network~\cite{DKT}. Dynamic Key-Value Memory Network (DKVMN) enhances the ability of the recurrent neural network by introducing two memory matrices to represent knowledge and learner's mastery for each knowledge respectively~\cite{DKVMN}. Graph-based Knowledge Tracing (GKT) combines knowledge tracing with a graph neural network~\cite{GKT}. It encodes a learner's hidden knowledge state as embeddings of graph nodes and updates the state in a knowledge graph. Those models have been proved effective but still have limitations. Most existing methods face the problem of exercise representation loss since they have not taken the texts of exercises into consideration. For the exercise based track, to the best of our knowledge, Exercise Enhanced Knowledge Tracing (EKT) is the first method to incorporate features of exercise text into the knowledge tracing model~\cite{EKT}. However, EKT extracts features of text by feeding the text of exercise directly into a bi-directional LSTM network~\cite{memeory}, which fails to consider the latent hierarchical graph nature of exercises and brings in extra noise from embeddings of text.

\begin{figure}[htb]
	\centering
	\includegraphics[width=\linewidth]{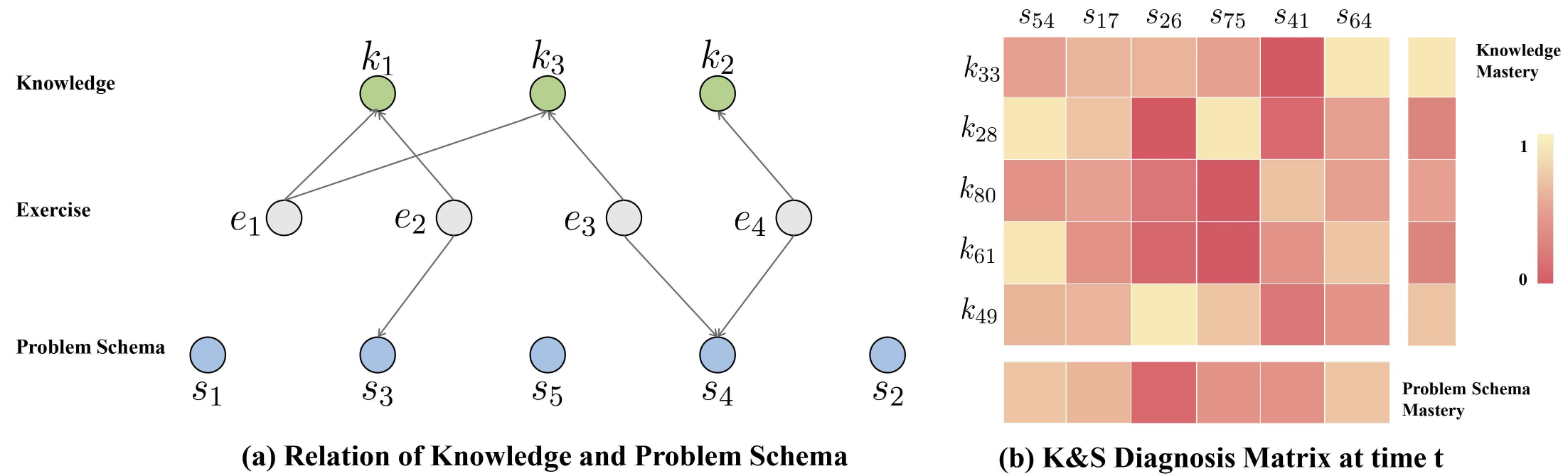}
	\caption{The left image shows the relations between knowledge and problem schema while the right image shows the diagnosis results of K\&S diagnosis matrix at time $\emph{t}$}
	\label{fig:relation}
\end{figure}

Our insight of solving the exercise representation loss problem and insufficient diagnosis problem hinges upon the idea of fully exploring latent hierarchical graph relations between exercises. Incorporating hierarchical relations between exercises could not only increase the accuracy of learner's performance prediction, but also enhance the interpretability of knowledge tracing. Fig.~\ref{fig:inspiration} clearly illustrates how hierarchical relations affect knowledge diagnosis results and the advantages of our approach compared to traditional knowledge tracing methods. As researchers have demonstrated the effectiveness of exploiting prerequisite relations into KT tasks\cite{chen2018prerequisite}, we decompose the hierarchical graph relations between exercises into direct support and indirect support relations. The intuition of support relations is that they represent different types of exercise learning dependencies, which could act as constraints for a knowledge tracing task. Moreover, to learn more fine-grained representation of exercises and avoid single exercise noises, inspired by math word solving technique~\cite{zhang2019gap,fuchs2004enhancing},  we introduce the concept of problem schema to summarize a group of similar exercises with similar solutions. The relation of two exercises is indirect support only if they belong to the same problem schema. It is worth mentioning that, we assume each exercise only has one main knowledge and one problem schema. Considering that exercises belonging to different knowledge could have similar solutions, and exercises with the same knowledge may also belong to different problem schemas due to their difficulty difference, we assume the relations between knowledge and problem schemas are many-to-many (as shown in Fig.~\ref{fig:relation}(a)).

The above analysis shows the prospects of introducing prior exercises support relations into a KT task. However, it may also bring in the following problems. First, the direct support relations between exercises can be defined in multiple ways, but which is the most suitable one for KT tasks remains unknown. Second, the definitions of problem schema and indirect support relations require understanding of exercises from a semantic perspective. How to automatically understand and represent the information is still challenging. Third, the hierarchical exercise relations contain different levels of features of exercises, how to organically combine these different levels of features is still worth exploring. Last, after we encode the information of hierarchical relations, we also hope that the model can always draw on past key information during current prediction. As shown in Fig.~\ref{fig:inspiration}, during making predictions for $e_{7}$, our model needs easily look back to important historical information such as the learner's wrong answer in $e_{2}$ or the relation between $e_{2}$ and $e_{7}$.

\iffalse
However, it may also bring in the following three problems. Firstly, the definition of the direct support and indirect support relations require lots of expert time, and manual labeled mappings are hard to generalize to other datasets. Secondly, it remains unexplored that what is the most effective way to map an exercise to a problem schema. Thirdly, the relationship between problem schemas and knowledge is complex. In particular, as Fig.~\ref{fig:first_concept} shows, two exercises (e.g., $e_{3}$ and $e_{4}$) with the same problem schema may contain different knowledge since they may contain multiple knowledge distribution. Here, we assume each exercise contains a distribution of knowledge while it has one major knowledge like most knowledge tracing methods.Similarly, two exercises (e.g., $e_{1}$ and $e_{2}$) with the same knowledge can also belong to different problem schemas due to their difficulty difference.
\fi

To address above challenges, we propose a hierarchical graph knowledge tracing framework named HGKT, which unifies the strengths of hierarchical graph neural network and recurrent sequence model with attention to enhance the performance of knowledge tracing. To summarize, the main contributions of this paper are as follows:

\iffalse
In HGKT, we present several data-driven methods to model direct support relations, and also introduce a way to semantically model the indirect support relations. Besides, we build a hierarchical exercise graph (HEG) consisting above two relations and leverage attention mechanisms to improve the accuracy of the KT task. To make a more detailed and convincing diagnosis results, a To summarize, the main contributions of this paper are as follows:
\fi

\begin{itemize}
	
	\item We introduce the concept of hierarchical exercise graph consisting of direct and indirect support relations between exercises, which could serve as learning constraints for a knowledge tracing task. We also present several data-driven methods to model direct support relations and introduce a way to semantically model the indirect support relations.
	
	\item We propose the concept of problem schema for exercises and explore a novel approach called hierarchical graph neural network to learn a refined representation for problem schema. This approach could help to address the exercise representation loss problem.

	\item We present two attention mechanisms that can highlight the important states of learners and fully leverage the information learned in HEG.
	
	\item To make the diagnosis results detailed and convincing, we put forward a knowledge\&schema (K\&S) diagnosis matrix  that can trace the mastery of both knowledge and problem schema at the same time (as shown in Fig.~\ref{fig:relation}(b)), which can also contribute to solving the insufficient diagnosis problem.
\end{itemize}

\begin{figure*}[htb]
	\centering
	\includegraphics[scale=0.415]{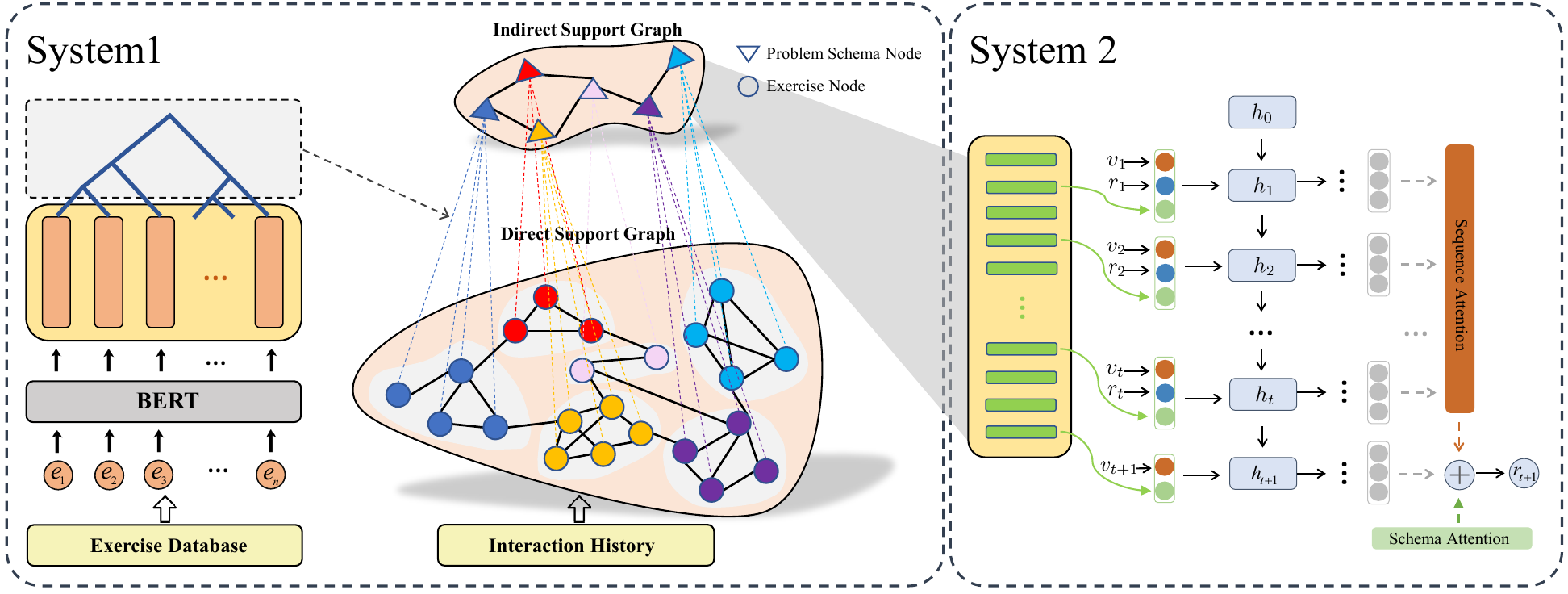}
	\caption{Overview of HGKT implementation. When predicting the performance of a leaner, System 1 generates problem schema embeddings fusing exercise hierarchical relations by HGNN and feed this information to System 2. System 2 incorporating graph and sequential information and updates learner's knowledge mastery at each time.}
	\label{fig:HGKT_framework}
\end{figure*}

\section{Problem Definition}
We formulate the KT task in this paper as follows. The goal of knowledge tracing is to estimate learners’ hidden knowledge states based on learners’ exercise data. Suppose we have a learner set $\mathcal P$, an exercise set $\mathcal E$, a knowledge set $\mathcal K$ and a problem schema set $\mathcal S$. Each exercise associates with a knowledge $k \in \mathcal K$ and a problem schema $s \in \mathcal S$. We further record exercise sequence for each learner $\emph{l}$ as $\mathcal R_{l} =\{(e_{1},r_{1}),(e_{2},r_{2})…(e_{m},r_{m})\}$, where $e_{i} \in \mathcal E$ and $r_{i}\in \{0,1\}$. Here $r_{i}=1$ indicates a correct answer and $r_{i}=0$ indicates an incorrect one. To clearly explain the framework of HGKT, we first briefly introduce the definition of HEG.

\subsubsection{Hierarchical Exercise Graph (HEG)} In an educational system, HEG is a hierarchical graph consisting of different levels of graphs that store prior direct and indirect support relations between exercises. Specifically, the HEG used in HGKT has two graphs : the bottom one is called direct support graph while the top one is called indirect support graph. Each node in the bottom graph corresponds to an exercise in our database while each node in top graph corresponds to a problem schema. The graph links in the bottom graph model the direct support relations while the relations between those two graphs model the indirect support relations. Formally, we represent a HEG graph as $(A,F,S_{e})$, where $A \in \{0,1\}^{\left|\mathcal E\right|*\left|\mathcal E\right|}$ is an adjacency matrix of the bottom graph, $F \in \mathbb{R}^{\left|\mathcal E\right|*t}$ is the node feature matrix assuming each node has $\emph{t}$ features and $S_{e} \in \mathbb{R}^{\left|\mathcal E\right|*\left|\mathcal S\right|}$ is an assignment matrix which denotes an assignment of each node from the bottom graph to a node in the top graph.

\iffalse
\newtheorem{ddd1}{Knowledge Tracing Problem}
\begin{definition}
	\textbf{Knowledge Tracing Task (KT Task)}. In an intelligent learning system, we assume that each learner contains a hidden temporal knowledge state for each concept at a time step, which denotes the mastery of each concept. Given an exercise interaction sequence for a learner $\emph{s}$ before time $\emph{t}$, the goal of a KT task is to find a method (i.e., train a model) to predict the learner's answer $r_{t+1}$ on the exercise $q_{t+1}$ at time $\emph{t+1}$. After the model has been trained, we could use it to infer the learner's hidden knowledge state at any time.
	
\end{definition}
\fi

\section{HGKT Framework}
\subsection{Framework Overview}
The entire structure of the framework is shown in Fig.~\ref{fig:HGKT_framework}. Here, System 1 aims to learn hierarchical graph structure of exercises by a hierarchical graph neural network (HGNN) and provides those resources to System 2.  System 2 then conducts the sequential processing, makes performance prediction and collects clues for System 1 to better refine the embeddings of each problem schema. To start training, we first need to generate the HEG from exercise database and interaction history. Thus, we introduce several methods to construct the direct support graph in section~\ref{section:s4.2} and to extract indirect support relations from a semantic perspective in section~\ref{section:psrl}. After the construction of HEG, exercising records and the problem schema embeddings are combined in a recurrent neural network (RNN) to predict the learner's performance. It is worth mentioning that HGNN and RNN are trained in an end-to-end fashion.

\iffalse
 First of all, many successes have demonstrated BERT's ability to understand the semantic information behind the text recently~\cite{Bert}, thus it is used to encode all exercises in the database to obtain the semantic representation. Next, we extract hierarchical graph relations from those embedding, which are combined to generate the hierarchical exercise graph. The third step is to learn the representation of problem schema by a hierarchical graph neural network (HGNN). Finally, the problem schema embedding and exercising records are used in a recurrent neural network (RNN) to predict the learner's performance. It is worth mentioning that the HGNN and RNN are trained in an end-to-end fashion.
\fi

\subsection{Direct Support Graph Construction}\label{section:s4.2}
HGKT can leverage prior hierarchical relations among exercises for a KT task as shown in Fig.~\ref{fig:inspiration}. However, the hierarchical relations of exercises are not explicitly provided in most cases. In this section, we first introduce several methods to explore the direct support relations among exercises, which are further used for learning the representations of problem schema. Many predecessors have proven the effectiveness of introducing graph structures into a KT task ~\cite{GKT,wang2019deep}, thus we propose several similarity rules based graph structures for exercises in section \ref{section:Knowledge-based}, \ref{section:Bert-Sim}, \ref{section:Exercise Transition}. Moreover, based on Bayesian statistical inference, we also present a method which could leverage the prior exercise support relations from the exercise interaction sequences in the training set and construct the direct support graph. To modeling the prior exercise relations as constraints, we first define the following properties about exercise support relations: 

\noindent{$\textbf{\emph Property\ 1}$}: We use $Sup(e_{1} \rightarrow e_{2})$ to denote the support degree of two exercises $e_{1}$ and $e_{2}$. $R_{e_{i}}$ and $W_{e_{i}}$ denote an event that whether a learner give right or wrong answer to $e_{i}$. A large value of $Sup(e_{1} \rightarrow e_{2})$ indicates the solutions of $e_{1}$ and $e_{2}$ have a strong support, which means if we know that a learner did $e_{1}$ wrong, then he has a high probability of doing $e_{2}$ wrong. In addition, if the learner is known to have done $e_{2}$ correctly, then the probability of him doing correct $e_{1}$ is also high.

\begin{align*}
P(R_{e_{1}}|R_{e_{2}}) > P(R_{e_{1}}|R_{e_{2}}, W_{e_{2}}),P(W_{e_{2}}|W_{e_{1}}) > P(W_{e_{2}}|R_{e_{1}}, W_{e_{1}}),\\
 if \ Sup(e_{1} \to e_{2})>0. \numberthis \label{eq:p1}
 \end{align*}  
 
\noindent{$\textbf{\emph Property\ 2}$}: Conversely, if $Sup(e_{1} \rightarrow e_{2})$ is small, it means that there is no prior support relations between the content and the solutions of the two exercises. In other words, the performance of learners for the two exercises is two unrelated events. Thus, we could conduct the following formulas:

\begin{align*}
P(R_{e_{1}}|R_{e_{2}}) = P(R_{e_{1}}|R_{e_{2}}, W_{e_{2}}),P(W_{e_{2}}|W_{e_{1}}) = P(W_{e_{2}}|R_{e_{1}}, W_{e_{1}}),\\
if \ Sup(e_{1} \to e_{2})=0. \numberthis \label{eq:p2}
\end{align*}

\iffalse
\begin{equation}
P(S{e_{j}} =0|\delta{(e_{i})=0}) > P(\delta{(e_{i})=1}), 

if \ Sup(e_{1} \textrightarrow e_{2}) > 0,
\end{equation}  
\fi

Based on the above reasoning, we construct the value of support between exercises as follows. Here,  $Count((e_{i},e_{j})=(r_{i},r_{j}))$ counts the number of exercise sequences that reply $e_{i}$ with an answer $r_{i}$ before $e_{j}$ with an answer $r_{j}$. Besides, to prevent the denominator from being too small, we introduced the laplacian smoothing parameter $\lambda_{p} = 0.01$~\cite{field1988laplacian} in Eq.(\ref{eq:p4}) and Eq.(\ref{eq:p41}).

\begin{equation}
P(R_{e_{1}}|R_{e_{2}}) = \frac{Count((e_{2},e_{1})=(1,1)) + \lambda_{p}}{\sum_{r_{1}=0}^{1}Count((e_{2},e_{1})=(1,r_{1})) + \lambda_{p}}, \numberthis \label{eq:p4}
\end{equation}

\begin{equation}
P(R_{e_{1}}|R_{e_{2}}, W_{e_{2}}) =\frac{\sum_{r_{2}=0}^{1}Count((e_{2},e_{1})=(r_{2},1)) + \lambda_{p}}{\sum_{r_{2}=0}^{1}{\sum_{r_{1}=0}^{1}}Count((e_{2},e_{1})=(r_{2},r_{1})) + \lambda_{p}}. \numberthis \label{eq:p41}
\end{equation}  

Similarly, we could also estimate the probability of $P(W_{e_{2}}|W_{e_{1}})$ and $P(W_{e_{2}}|R_{e_{1}}, W_{e_{1}})$. The value of support is defined as the sum of the following two components. Here, $max $ function is used to ensure the non-negativity of support value.

\iffalse
\begin{equation}
\frac{P(R_{e_{1}}|R_{e_{2}})}{P(R_{e_{1}})} = \frac{Count((e_{2},e_{1})=(1,1)) + \lambda}{\sum_{r_{1} \in \{0,1\}}Count((e_{2},e_{1})=(1,r_{1})) + \lambda} \\
, \numberthis \label{eq:p4}
\end{equation}

\begin{equation}
\frac{P(W_{e_{2}}|W_{e_{1}})}{P(W_{e_{2}})} = \frac{Count((e_{1},e_{2})=(0,0)) + \lambda}{\sum_{r_{2} \in \{0,1\}}Count((e_{1},e_{2})=(0,r_{2})) + \lambda} \\
   , \numberthis \label{eq:p5}
\end{equation}    
\fi

\begin{align*}
Sup(e_{1} \to e_{2}) = max(0, \ln{\frac{P(R_{e_{1}}|R_{e_{2}})}{P(R_{e_{1}}|R_{e_{2}},W_{e_{2}})}}) \ \\ 
+\  max(0,\ln{\frac{P(W_{e_{2}}|W_{e_{1}})}{P(W_{e_{2}}|R_{e_{1}}, W_{e_{1}})}}).
\numberthis \label{eq:p3}
\end{align*}

\subsubsection{Knowledge-based Method}\label{section:Knowledge-based} generates a densely connected graph, where $A_{i,j}$ is \emph{1} if two different exercises $e_{i}$ and $e_{j}$ contain same knowledge; else 
it is \emph{0}.

\subsubsection{Bert-Sim Method}\label{section:Bert-Sim} generates a graph by the cosine similarity of two exercises' BERT embedding. Here $A_{i,j}$ is \emph{1} if the similarity between two different exercises is larger than hyperparameter $\omega$; else, it is \emph{0}.

\subsubsection{Exercise Transition Method}\label{section:Exercise Transition} generates a graph whose adjacency matrix is a transition probability matrix, where $A_{i,j}$ is \emph{1} if $\frac{n_{i,j}}{\sum_{k=1}^{\left|\mathcal E\right|}n_{i,k}} > \omega$; else, it is \emph{0}. Here $n_{i,j}$ represents the number of times exercise $\emph{j}$ is answered immediately after exercise $\emph{i}$ is answered.

\subsubsection{Exercise Support Method}\label{section:Exercise Support} generates a graph
by bayesian statistical inference, where $A_{i,j}$ is \emph{1} if $Sup(e_{i},e_{j})
> \omega$; else, it is \emph{0}.

\subsection{Problem Schema Representation Learning} \label{section:psrl}
In this section, we first describe methods to explore the indirect support relations between exercises. The hierarchical relations extracted are used to compose the HEG. The goal of System 1 is to learn representations of problem schema for each exercise, thus we also propose a way to fuse those hierarchical relations.

The essence of mining indirect support relations is to find the corresponding problem schema for each exercise (as shown in Fig.~\ref{fig:inspiration}), which can be transformed into a problem of unsupervised clustering of exercises. Considering the semantic nature of problem schema, we use BERT~\cite{Bert} to encode all exercises in the database to obtain its semantic representation as many successes have demonstrated BERT's ability to understand the semantic information behind texts. Besides, to better obtain the clustering results with multiple levels to accommodate the cognitive nature of students at different levels, we adopt the hierarchical clustering~\cite{johnson1967hierarchical} to cluster the BERT embeddings of exercises.  Hierarchical clustering is an unsupervised cluster analysis method using agglomerative or divisive strategies to build a hierarchy of clusters. We could set different clustering threshold $\lambda$ to get different levels of clustering results, which could be used to find the most suitable level for problem schema. Moreover, in order to better combine the graph structures and the clustering results of the exercises, motivated by the assignment matrix proposed in DiffPool~\cite{ying2018hierarchical}, we propose an exercise assignment matrix $S_{e}$ that can provide an assignment of each exercise node in the direct support graph to a problem schema node in the indirect support graph. Here, each row of the $S_{e}$ corresponds to one of exercise at bottom graph, and each column of $S_{e}$ corresponds to one of problem schema at top graph. It is worth mentioning that the assignment matrix in DiffPool is learned by a separate GNN model which is computational expensive and hard to control hierarchical relations in our cases. Thus, we use hierarchical clustering to generate the assignment matrix to store indirect support relations.

\begin{figure}[htb]
	\centering
	\includegraphics[width=1.0\linewidth]{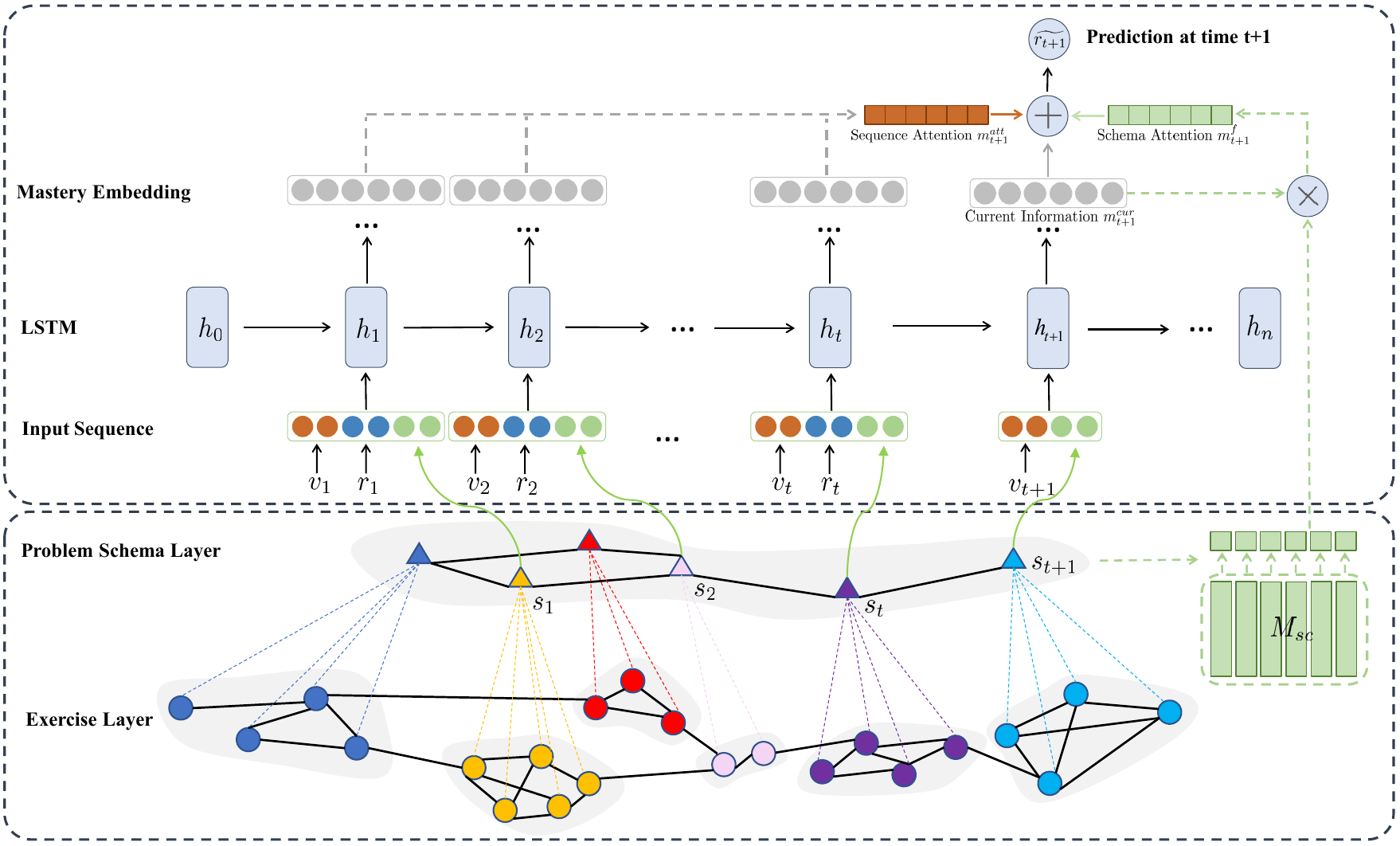}
	\caption{The architecture of HGKT}
	\label{fig:HGKT_architectures}
\end{figure}

After extracting the hierarchical relations from exercises, we now introduce the detailed strategie of fusing graph information in HEG. Here we propose a hierarchical graph neural networks (HGNN) to propagate semantic features of exercises into problem schema embedding in the HEG (as shown in Fig.~\ref{fig:HGKT_architectures}). The HGNN consists of convolutional layers and pooling layers~\cite{kipf2016semi,lee2019self,scarselli2008graph}. The key intuition of HGNN is that we stack a direct support graph with exercise information and a indirect support graph with problem schema information, and leverage assignment matrix to assign each node from direct support graph to indirect support graph. Formally, for a given HEG = $(A,F,S_{\lambda})$, the HGNN propagates features using following equations. First of all, we build two GNN networks named $GNN_{exer}$ and $GNN_{sche}$ whose parameters are used to update exercise embedding and problem schema embedding correspondingly. The node feature matrix for the first $\emph{k}$ layers is one-hot embedding of each exercise $F$ and a direct support graph $A_{e}$. Here $H_{e}$ and $H_{s}$ correspond to the embedding of exercises and problem schemas. Note that $H^{0}=F$. At $\emph{k-th}$ layer, as shown in Eq.~\ref{eq:aa}, we utilize a pool operation to coarsen the direct support graph to get a smaller indirect support graph $A_{s}$. The linear transformation introduced in Eq.~\ref{eq:sh} aggregates the exercise representations to correspond problem schema representations. At last, $GNN_{sche}$ updates each embedding of problem schema and sends this information to sequence processing stage of HGKT.

\begin{equation}
H_{e}^{(l+1)} = GNN_{exer}(A_{e},H_{e}^{(l)}), l<k,
\label{eq:exer}
\end{equation}  

\begin{equation}
A_{s} = S_{\lambda}^\mathrm{T}A_{e}S_{\lambda}, l=k,
\label{eq:aa}
\end{equation}  

\begin{equation}
H_{s}^{(l+1)} = S_{\lambda}^\mathrm{T}H_{e}^{(l)}, l=k,
\label{eq:sh}
\end{equation}

\begin{equation}
H_{s}^{(l+1)} = GNN_{sche}(A_{s},H_{s}^{(l)}), l>k.
\label{eq:sche}
\end{equation}

\subsection{Sequence Modeling Process}
This section mainly focuses on combining the representations of problem schemas 
with sequence information.

\subsubsection{\textbf{Sequence Propagation}} The overall input for the sequence processing part is exercise interaction sequences. Each exercise interaction contains three components : knowledge $v_{t}$, exercise result $r_{t}$, problem schema $s_{t}$. Here, $v_{t}$ is a one-hot embedding of $\left|\mathcal K\right|$ distinct knowledge. $r_{t}$ is a binary value indicating whether the learner answers an exercise correctly. $s_{t}$ is an embedding of problem schema generated in HEG. At each time step $\emph{t}$, to differentiate the exercise contribution to its corresponding knowledge hidden state, the input for the sequence model is a joint embedding $x_{t}$ of $(v_{t},r_{t},s_{t})$. In the propagation stage, we process $x_{t+1}$ and the previous hidden state $h_{t}$ using RNN network to get current learner's hidden state $h_{t+1}$ as shown in Eq.~\ref{lstm}. Here we use LSTM as a variant of RNN since it can better preserve long-term dependency in the exercise sequences~\cite{memeory}. Eq.~\ref{cur} shows the prediction of mastery for each problem schema $m_{t+1}^{cur} \in \mathbb{R}^{\left|\mathcal S\right|}$ at time \emph{t+1}. $\{W_{1},b_{1}\}$ are the parameters.

\begin{equation}{\label{lstm}}
h_{t+1},c_{t+1} = LSTM(x_{t+1},h_{t},c_{t};\theta_{t+1}),
\end{equation}  

\begin{equation}{\label{cur}}
m_{t+1}^{cur}= ReLU(W_{1} \cdot h_{t+1}+b_{1}).
\end{equation}

\subsubsection{\textbf{Attention Mechanism}} The HGKT leverages two kinds of attention mechanism, called sequence attention and schema attention, to enhance effects of typical states in the history of modeling. Learners may perform similarly on exercises with same problem schemas, thus we use a sequence attention shown in Eq.~\ref{seqa} to refer to previous results of similar exercises. Here, we assume the attention of historical problem schema mastery $m_{t+1}^{att}$ is a weighted sum aggregation of previous mastery state. However, different from the attention used in ~\cite{EKT}, we set an attention window limit $\lambda_{\beta}$ in HGKT for the following two reasons: (1) If the length of sequence attention is not limited, the computational cost would be intensive when the exercise sequence is extremely long. (2) Experiments prove that recent memory has a greater impact on knowledge tracing results than long past memories, which is consistent with educational psychology since learners start losing the memory of learned knowledge over time~\cite{ebbinghaus2013memory}.

\begin{equation}{\label{seqa}}
m_{t+1}^{att}= \sum_{i=max(t-\lambda_{\beta},0)}^{t}\beta_{i}m_{i}^{cur}, 
\beta_{i}=\cos(s_{t+1},s_{i}).
\end{equation} 

\iffalse
\begin{figure*}[htb]
	\centering
	\includegraphics[scale=0.5]{figures/test_stage}
	\caption{The illustration of testing stage and K\&S diagnosis matrix}
	\label{fig:test_stage}
\end{figure*}
\fi

The schema attention aims to focus learner's attention on a given problem schema with $\alpha_{t} \in \mathbb{R}^{\left|\mathcal S\right|}$, which is the similarity to other problem schemas. As shown in Eq.~\ref{eq:schea}, we propose an external memory $M_{sc} \in \mathbb{R}^{k*\left|\mathcal S\right|}$ collected from embeddings in the final layer of indirect support graph. Each column of $M_{sc}$ corresponds to an embedding of problem schema. Here, \emph{k} is the embedding dimension in HEG. The intuition for Eq.~\ref{eq:schea} is that the answer of an exercise would relate to exercises with similar problem schema, thus we could focus our attention on a certain problem schema. Notice that the value of memory $M_{sc}$ changes over time during training.

\begin{equation}
m_{t+1}^{f} = \alpha_{t+1}^{T}m_{t+1}^{cur}, 
\alpha_{t+1}=Softmax(s_{t+1}^TM_{sc}).
\label{eq:schea}
\end{equation}

To summarize, the states to predict the learner's performance at time \emph{t+1} consists of three components : current knowledge mastery $m_{t+1}^{cur}$, historical related knowledge mastery $m_{t+1}^{att}$ and focus problem schema mastery $m_{t+1}^{f}$. As Eq.~\ref{eq:pred} shows, those states are concatenated together to predict a final result. $\{W_{2},b_{2}\}$ are the parameters.

\begin{equation}
\widetilde{r_{t+1}}= \sigma(W_{2} \cdot 
[m_{t+1}^{att},m_{t+1}^{cur},m_{t+1}^{f}]+b_{2}).
\label{eq:pred}
\end{equation}

\subsubsection{\textbf{Model Learning}} The goal of training is the negative log likelihood of the observed sequence of learner response. During training, both parameters, i.e., parameters in $GNN_{exer}$ and $GNN_{sche}$, parameters in sequence propagation $\{W_{1},b_{1},W_{2},b_{2}\}$ are updated. The loss for a learner is shown in Eq.~\ref{eq:loss}. Specifically, the loss for a response log is defined as the cross entropy between the real answer $r_{t}$ at time $\emph{t}$ and predicted score $\widetilde{r_{t}}$ on exercise. The objective function is minimized using the Adam optimization~\cite{kingma2014adam}. More implementation details will be introduced in the experiments section.

\begin{equation}
loss = 
-\sum_{t}(r_{t}log\widetilde{r_{t}}+(1-r_{t})log(1-\widetilde{r_{t}})).
\label{eq:loss}
\end{equation}

\subsection{Prediction Output of HGKT}
After modeling the exercising process of each learner from step \emph{1} to \emph{t}, we now introduce our strategies for predicting his performance on the next exercise $e_{t+1}$. Moreover, to diagnose a learner's learning process in detail, we introduce a K\&S diagnosis matrix to fully trace the mastery of both knowledge and problem schema dynamically.

\subsubsection{\textbf{K\&S Diagnosis Matrix}}
Unlike the traditional knowledge tracing method, the input for HGKT in testing stage is the knowledge and problem schema of an exercise. Thus, we could trace the transition mastery of knowledge, problem schema, or their combination in HGKT. Specifically, at each time step $\emph{t}$, we can predict the performance $r_{t}\label{key}$ for each combination of $(k_{i},s_{j})$ ($k_{i} \in \mathcal K, s_{j} \in \mathcal S$). Thus, we could use those results to generate a matrix $R_{t}^{ks}$ called K\&S diagnosis matrix whose vertical axis represents different knowledge and horizontal axis represents different problem schema. Next, the knowledge mastery $R_{t}^k$ at time $\emph{t}$ is calculated by the weighted aggregation of mastery of each problem schema as shown in Eq.~\ref{eq:d1}. Here $q_{i,j}$ denotes the number of exercises containing knowledge $k_{i}$ and problem schema $s_{j}$. Similarly, we could calculate the mastery of each problem schema $R_{t}^s$.

\begin{equation}
q_{i,j} = \left|{\{(e_{(k_{i},s_{j})}\mid k_{i}\in \mathcal K,s_{j}\in \mathcal 
	S)}\}\right|
\label{eq:d}
\end{equation}

\begin{equation}
R_{t,i}^k = R_{t,i}^{ks}d_{i}^{k},\ d_{i,j}^{k} = 
\frac{q_{i,j}}{\sum_{j}q_{i,j}} 
\label{eq:d1}
\end{equation}

\begin{equation}
R_{t,j}^s = R_{t,j}^{ks}d_{j}^{s},\ d_{i,j}^{s} = 
\frac{q_{i,j}}{\sum_{i}q_{i,j}}. 
\label{eq:d2}
\end{equation}

\subsubsection{\textbf{Interpretability of Problem Schema}} The introduction of problem schema can effectively improve the effect of models in predicting learners' performance. However, the interpretability of problem schema is unknown. Based on above challenges, we propose an unsupervised schema summarization algorithm. The core idea of this algorithm is to utilize TextRank~\cite{textrank} to extract meaningful condition descriptions and objective descriptions accordingly and use them to form a description for an exercise cluster based on certain rules. Table~\ref{tab:Summarization} shows an example of the summarization of a group of exercises. More details about the algorithm would be introduced in Appendix.

\section{Experiments}
In this section, we evaluate the performance of the proposed framework HGKT from following three aspects: (1) The prediction performance of HGKT against baselines. (2) The analysis of effectiveness of hierarchical exercise graph. (3) The effectiveness of sequence attention and schema attention mechanism used in HGKT.

\subsection{Experiment Setup}

\subsubsection{\textbf{Dataset and Preprocessing}} Since there is no open dataset which could provide exercising records with text information, our experimental dataset is derived from a large real-world online education system. The Aixuexi online system \footnote{www.aixuexi.com} could track exercising records of learners. Each record of a learner includes learner id, lecture id, class type, question information, answer time, learners' answers and correction results. The question information includes the question id, question content, question type and the primary knowledge contained in the question. To avoid data distribution differences, we use the data obtained after 2018. In the data preprocessing stage, we group the exercising records with learner id and sort the records according to the response time. In total, we get 91,449,914 answer records from 132,179 learners. The detailed statistics of dataset are shown in Table~\ref{tab:statistic}.

\iffalse
\begin{figure}[htb]%
	\centering
	\subfloat[Distribution of exercise records 
	]{{\includegraphics[width=3.5cm]{number_of_exercising_records}}}%
	\qquad
	\subfloat[Distribution of content length
	]{{\includegraphics[width=3.5cm]{content_length}}}%
	\caption{The distribution of Aixuexi dataset}% 
	\label{fig:distribution}%
\end{figure}
\fi

\begin{figure*}[htb]
	\centering
	\includegraphics[width=\linewidth]{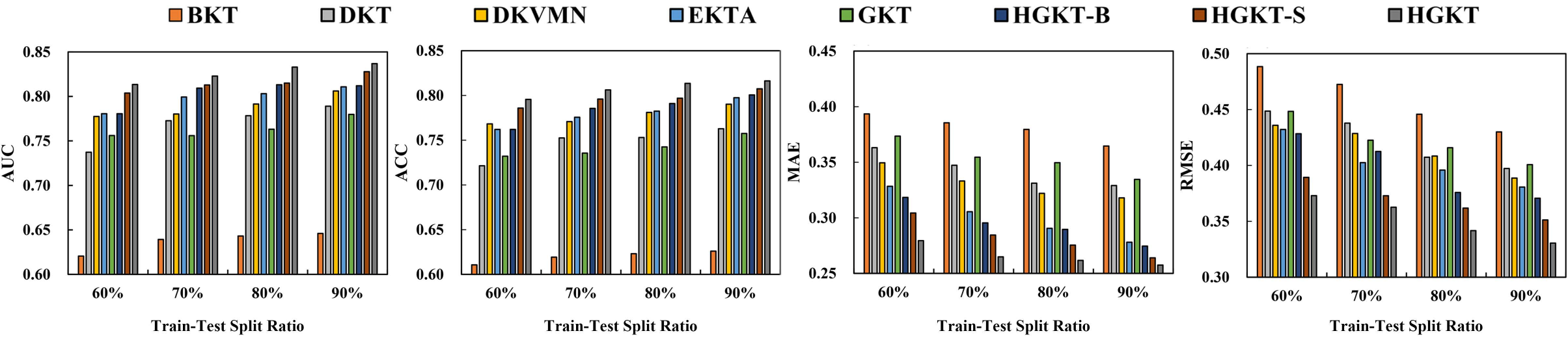}
	\caption{The results of learner performance prediction under four metrics.}
	\label{fig:all_model_auc}
\end{figure*}

\renewcommand{\algorithmicrequire}{\textbf{Input:}} % Use Input in the format 
%of Algorithm
\renewcommand{\algorithmicensure}{\textbf{Output:}} % Use Output in the format 
%of Algorithm

\begin{table}[]
	
	\caption{An Example of Schema Summarization}
	\centering
	\resizebox{0.45\textwidth}{!}{%
		\begin{tabular}{|p{9cm}|}
			\hline
			\textbf{Exercise   Cluster} \\ \hline
			\emph{Exercise 1}: If the ratio of lengths of three sides of a triangle is 
			$2:3:4$, and its circumference is 18, the shortest side length is?  
			\\ 
			\emph{Exercise 2}: Given ratio of lengths of triangle sides is 2:4:4 and 
			circumference is 20, what is the shortest side length? \\ 
			\emph{Exercise 3}: If we know that ratio of lengths of three sides of a 
			triangle is 3:4:5, and circumference of the triangle is 24, find 
			the shortest side length? \\ 
			... \\ 
			\hline
			\textbf{Condition keyphrases} \\ 
			\hline
			\multicolumn{1}{|c|}{ratio of lengths, circumference} \\ 
			\hline
			\textbf{Objective keyphrases} \\ \hline
			\multicolumn{1}{|c|}{shortest side length} \\ \hline
			\textbf{Problem Schema} \\ \hline
			\multicolumn{1}{|c|}{Given ratio of lengths, circumference, find 
				the shortest side length?} \\ \hline
			
		\end{tabular}%
		\label{tab:Summarization}
	}
\end{table}

\begin{table}
	\centering
	\caption{The statistics of the Aixuexi dataset.}
	\label{tab:statistic}
	\begin{tabular}{c|c}
		\hline
		Statistics & Value\\
		\hline
		{number of learners} & {$132,179$}\\
		number of response logs & {$9,144,914$}\\
		{number of exercises correctly answered} & 
		{$6,747,435$}\\
		average of exercises in response log & {$69$}\\
		number of different exercises& {$11,410$}\\
		number of knowledge related& {$490$}\\
		total number of knowledge & {$672$}\\
		\hline
	\end{tabular}
	
\end{table}

\iffalse
\subsubsection{\textbf{Implementation Details}} We conduct extensive experiments to find the best parameters for HGKT models. To extract the semantic representation for exercises, we utilize a public BERT-Base tool~\cite{Bert} to convert each exercise into an embedding vector with 768 dimensions. In problem schema representation learning part, we use transition graph to construct the exercise graph. Besides, the best graph structure for HEG is three graph convolutional layers followed by a graph pooling layer and a graph convolutional layer. The size of graph hidden embedding is 64, the size of problem schema embedding is 30. The most suitable clustering threshold for HEG is 9 and we get 1136 problem schemas. In the sequence propagate stage, we set the attention window size to 20.
\fi

\subsubsection{\textbf{Implementation Details}} To set up the training of HGKT, we first need to use unsupervised data to generate the HEG. We leverage the exercising interaction data in the training set to construct the direct support graph and all exercise data involved in our experiment to construct the indirect support relations. Because the construction of the former one requires the learner's exercising results while the latter one only needs the exercise information. Specifically, we utilize a public BERT-Base tool \footnote{https://github.com/hanxiao/bert-as-service} without any fine-tune to convert each exercise into an embedding vector with 768 dimensions, and the hierarchical clustering tool \footnote{https://docs.scipy.org/doc/scipy/reference/cluster.hierarchy.html} during the indirect support relations construction process. The framework of HGKT is joint training after the construction of HEG. During the testing stage, we use the same HEG to make predictions.

We conduct extensive experiments to find the best parameters for HGKT. During HEG construction process, we use the exercise support method to construct a direct support graph and set the clustering threshold to 9 then get 1136 problem schemas to construct the indirect support graph. In HGNN, we get best results with three graph convolutional layers in $GNN_{exer}$ and one graph convolutional layer in $GNN_{sche}$. The exercise embedding size in HGNN is set to 64 and schema embedding is set to 30. In the sequence propagate stage, we set the attention window size to 20 and LSTM hidden embedding size to 200. During training, we use Adam optimizer with the learning rate 0.01 and set mini-batches as 32. We also use dropout \cite{srivastava2014dropout} with probability 0.5 to prevent overfitting. 

\iffalse
We will publish our code after the acceptance of our paper.
\fi

\subsubsection{\textbf{Comparison Baselines}} \label{sec:baselines} To demonstrate the effectiveness of our framework, we compare HGKT with following state-of-art methods. Those methods are chosen from three aspects: (1) Traditional educational models: Bayesian Knowledge Tracing (BKT) (2) Deep learning models:  Deep Knowledge Tracing (DKT), Dynamic Key-Value Memory Networks (DKVMN), Exercise-aware Knowledge Tracing (EKT), Graph Knowledge Tracing (GKT) (3) Variants of HGKT : HGKT-B, HGKT-S

\begin{itemize}	
	\item \textbf{BKT} is a traditional knowledge tracing model which is based 
	on
	hidden Markov model. The knowledge states for each concept are a set of 
	binary variables~\cite{BKT}.
	
	\item \textbf{DKT} uses a recurrent neural network 
	to model the learners'exercising process~\cite{DKT}. It models 
	question sequences as knowledge sequences. We follow the hyperparameters 
	mentioned in ~\cite{DKT}, where the size of the hidden layer is 200 and we 
	use a GRU for the RNN.
	
	\item \textbf{DKVMN} is a deep learning
	method using a key matrix to store knowledge
	representation and a value matrix for each learner to
	update the mastery for each concept~\cite{DKVMN}. In the experiment, 
	we get best results when we set the size of the memory slot is 30, the 
	size of embedding of key matrix is 50 and the size of embedding of value 
	matrix is 200.
	
	\item \textbf{GKT} is a GNN-based knowledge tracing method which 
	reformulate 
	the knowledge tracing task as a time-series node-level classification 
	problem in the GNN~\cite{GKT}. In experiments, we get the best results 
	using the 
	knowledge graph generated by transition-graph method.
	
	\item \textbf{EKTA} uses a bidirectional 
	LSTM to encode exercise text and leverage an attention mechanism to enhance 
	the prediction accuracy~\cite{EKT}. In experiments, we get best results when 
	the shape of its hidden state is 672 and 30.
	
	\item \textbf{HGKT-B} is a variant of HGKT framework. Here, in the modeling 
	process, we do not use problem schema embedding generated by HEG. Instead, 
	we use BERT to encode text features.
	
	\item \textbf{HGKT-S} is another variant of HGKT framework. In the modeling 
	process, we use the 
	one-hot of problem schema generated by hierarchical clustering to replace 
	embedding of problem schema generated by HEG.
\end{itemize}

\subsubsection{\textbf{Evaluation Setting}} We first compare the overall performance of HGKT with baseline models. In the experiment, we randomly select $60\%$,$70\%$,$80\%$,$90\%$ of the exercise records as training data, and the remaining as testing data. To qualify the effectiveness of the model, we compare the results of experiments from both regression and classification perspectives~\cite{fogarty2005case,kuang2018stable,wu2015cognitive}. We use open-source platform to implement most comparison baseline models and searched the hyperparameters to find the best performance for each model. Our framework for HGKT is implemented by Torch on a Linux server with eight Intel Xeon Skylake 6133 (2.5 GHz) CPUs and four Tesla V100 GPUs. For all models, we test different parameters and experiments five times, respectively, and finally take the average of metrics as evaluation metrics.

\subsection{Comparison}
Fig.~\ref{fig:all_model_auc} shows the overall comparing results on this task. The results indicate that HGKT performs better than other baseline models. Besides, we draw several conclusions from the results: (1) All our proposed HGKT based models perform better than other baseline models. The result clearly demonstrates that HGKT framework could make full use of the information contained in exercising records and exercise texts, benefiting the prediction performance. (2) The result of HGKT-B outperforms EKTA, which shows that BERT could better extract the representation of text than a Bi-LSTM neural network. (3) The HGKT-S has better performance than HGKT-B, which proves that compared to directly using the text embedding of each exercise, summarizing a bunch of similar exercises with problem schemas would bring in less noise. (4) The HGKT performs better than HGKT-S, which reveals that the introduction of HEG for problem schema representation learning is also effective. The reason may be that in HEG, the representation of each exercise is propagated through a HGNN, so that problem schema embedding could aggregate the information of a certain kind of exercises.

\subsection{Analysis} 

To understand the effects of the various components and hyperparameters in HGKT, we conduct ablation studies and parameter analysis on the Aixuexi dataset.

\subsubsection{\textbf{Ablation Study}} We investigate the effects of different support relations and attention mechanisms in our model. The results of ablation study are shown in Table \ref{tab:ablation_study}. To conduct a fair comparison experiment, we propose methods that have the same framework but have different components with HGKT. The first method in Table \ref{tab:ablation_study} adopts the same framework with HGKT but has no problem schema embeddings learning part. The second method does not contain hierarchical information of exercises and only leverages the bottom layer of HEG to learn the embeddings of each exercise. The third method only uses the information of indirect support relations, which is the one-hot of hierarchical clustering results with a dense layer to learn embeddings of each problem schema. The fourth method drops the hierarchical graph neural network part and directly merges the information learned by previous two methods. All of above methods have no attention component and those learned embeddings are used to replace the embeddings of problem schemas in HGKT framework. The fifth method only lacks the attention components compared to the HGKT. From the result, we observe that HGKT with both support relations and attention mechanisms performs the best. We also note that having direct support relations and indirect support relations still outperform the implementation without either relations. More interestingly, we also find that merging two support relations directly performs worse than using hierarchical graph neural network. We postulate that a possible reason for the inferior performance may be due to the extra noise introduced by single exercise. The HGNN part in HGKT acts as a coarse-grained structure extractor which also averages the features of a group of similar exercises, thus reduces the noise of exercises.

\begin{table}[]
	\caption{Model performance over different graph structure in HGKT. $\emph{GS}$ denotes the structure of graph. $B-i\_T-j$ indicates \emph{i}  graph convolutional layers in $GNN_{exer}$ and \emph{j}  graph convolutional layers in $GNN_{sche}$.}\label{tab:graph_number}
	\begin{tabular}{l|cc|l|cc}
		\hline
		GS & AUC (\%) & ACC (\%) &  GS & AUC (\%) & ACC (\%) \\ \hline
		B-1\_T-1 & 81.4 & 81.5 & B-2\_T-2 & 82.5 & 82.4 \\
		B-1\_T-2 & 82.3 & 82.4 & B-2\_T-3 & 81.7 & 81.6 \\
		B-1\_T-3 & 81.8 & 81.9 & B-3\_T-1 & \textBF{82.8} & \textBF{82.6} \\
		B-2\_T-1 & 82.4 & 82.4 & B-3\_T-2 & 82.4 & 82.4 \\ \hline
	\end{tabular}
\end{table}

\begin{table}[]
	\caption{Ablation study of HGKT}\label{tab:ablation_study}
	\begin{tabular}{cc|cc}
		\hline
		methods                         & attention & AUC (\%)  & ACC (\%)  \\ \hline
		no support relations	             & no        & 78.8 & 78.9 \\
		only direct support             & no        & 79.3 & 79.5 \\
		only indirect support           & no        & 80.4 & 80.2 \\
		merge two support relations          & no        & 80.3 & 80.5 \\
		both support relations          & no        & 81.2 & 81.1 \\ \hline
		(HGKT) both support   relations & yes       & \textBF{82.8} & \textBF{82.6} \\ \hline
	\end{tabular}
\end{table}

\subsubsection{\textbf{Graph Structure Analysis}} In HGKT, there are three factors that may affect the structure of hierarchical exercise graph: different direct support graph construction methods,  the problem schema clustering level and the number of layers used in HGNN.

\begin{figure}[htb]%
	\centering
	\subfloat[AUC between Graph 
	Methods]{{\includegraphics[width=3.5cm]{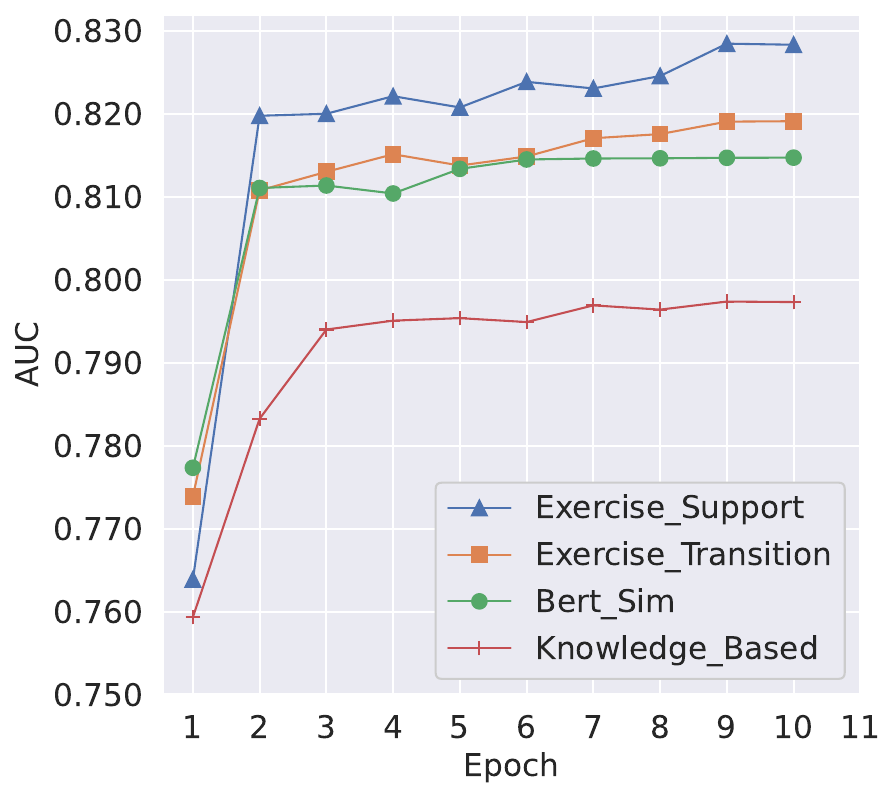} }}%
	\qquad
	\subfloat[AUC between edge-to-node
	Ratio]{{\includegraphics[width=3.5cm]{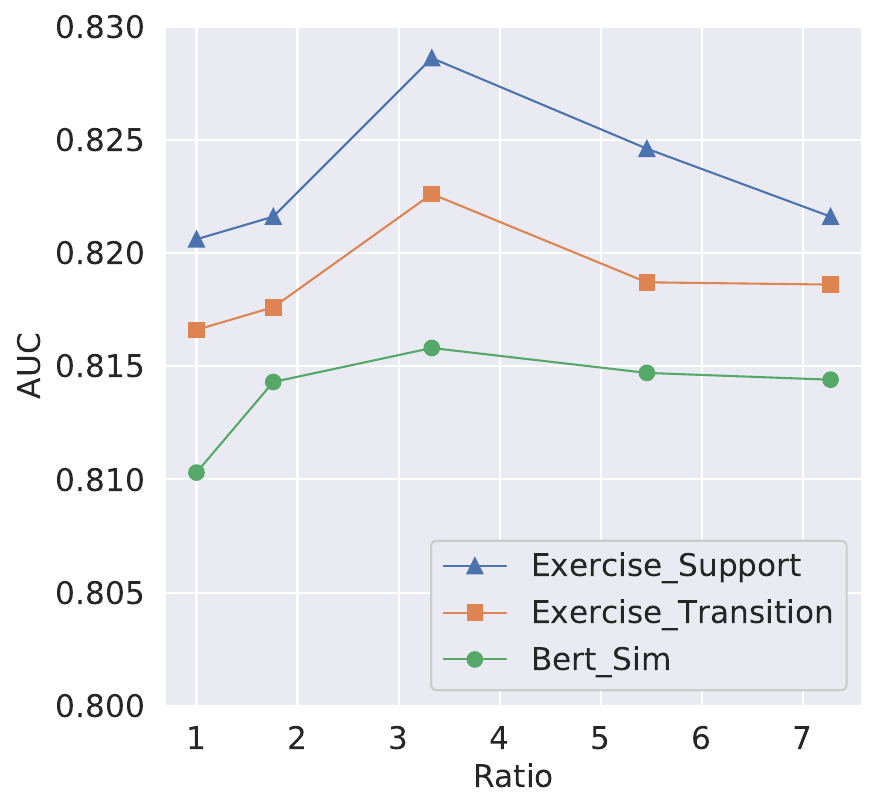} }}%
	\caption{The effectiveness of different graph methods}% 
	\label{fig:graph_auc}%
\end{figure}

In section~\ref{section:s4.2}, we propose several methods to explore different direct support relations of exercises. Those methods decide different graph structures of direct support graph in HEG. Fig.~\ref{fig:graph_auc}(a) shows the knowledge tracing AUC of different methods. From the figure, we find that bayesian-based graph construction method outperforms other methods. A potential reason could be due to that the bayesian-based method could employ the information in exercise interactions history rather just use the exercise features like other methods. We also test impacts of different edge-to-node ratio for different Graph (as shown in Fig.~\ref{fig:graph_auc}(b)). The result shows the effect of graph convolution is the best when the ratio of edge-to-node is about \emph{3}-\emph{4}.

\begin{figure}[htb]%
	\centering
	\subfloat[The Statistics of Number of Problem Schema 
	]{{\includegraphics[width=3.5cm]{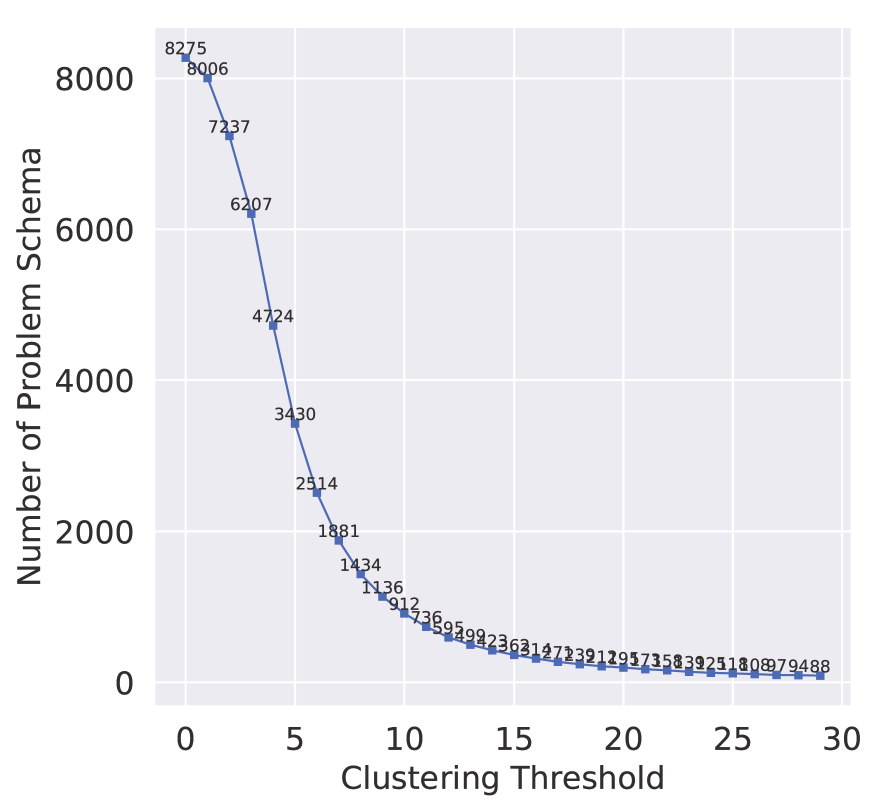}}}%
	\qquad
	\subfloat[AUC between Different Threshold 
	]{{\includegraphics[width=3.5cm]{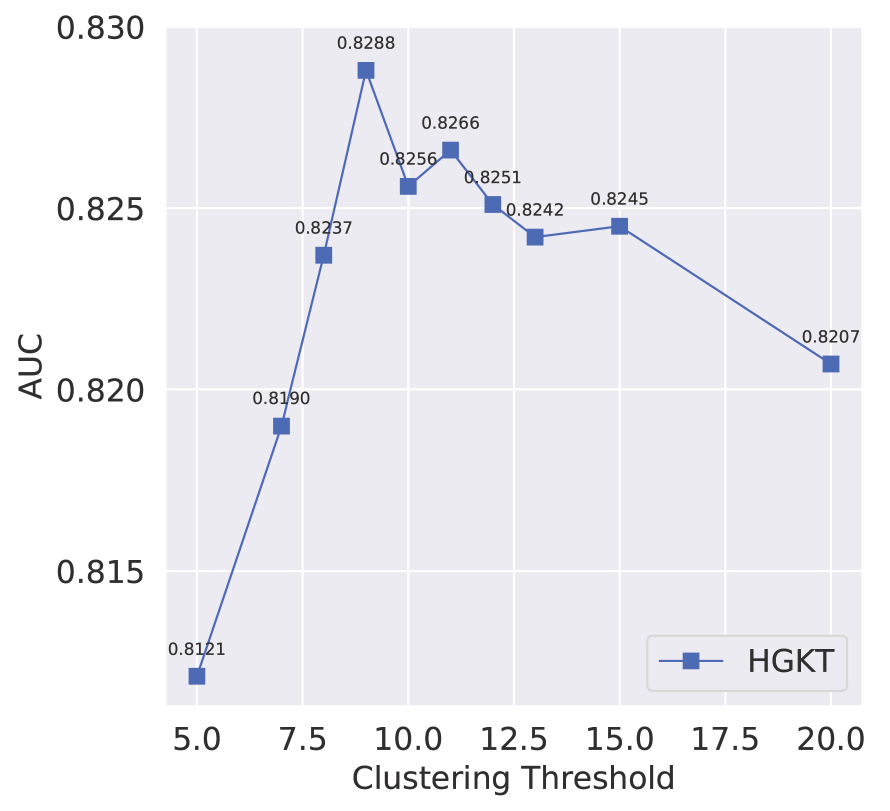}}}%
	\caption{Performance over different clustering values in HGKT}% 
	\label{fig:level_auc}%
\end{figure}

The problem schema clustering level affects the number of nodes in an indirect support graph. As shown in Fig.~\ref{fig:level_auc}(a), the number of problem schemas is more than \emph{3430} when the threshold is \emph{5}, which results in intensive computational cost, and the generated problem schemas are not typical enough to represent a group of exercises. Thus, we test the overall performance of HGKT with thresholds between \emph{5} to \emph{20}. Fig.~\ref{fig:level_auc}(b)   shows that the best AUC occurs when clustering threshold is 9 and the number of problem schema is 1136. Besides, from the overall trend of the curve, we could infer that there is a most suitable division of exercises for KT task, which means that using problem schema to represent a group of similar exercises is reasonable.

The numbers of graph convolutional layers used in the $GNN_{exer}$ and $GNN_{sche}$ are also tuneable hyperparameters. Thus, we investigate the effect of the numbers of graph convolutional layers on our model’s performance. Table \ref{tab:graph_number} shows the result of experiments. We find that $B-3\_T-1$ version achieves the best result for the KT task. The reason may be that the  $B-3\_T-1$ structure could achieve the optimal capacity of information aggregation.

\begin{figure}[!ht]
	\centering
	\subfloat[AUC between different Attention Methods 
	]{{\includegraphics[width=3.5cm]{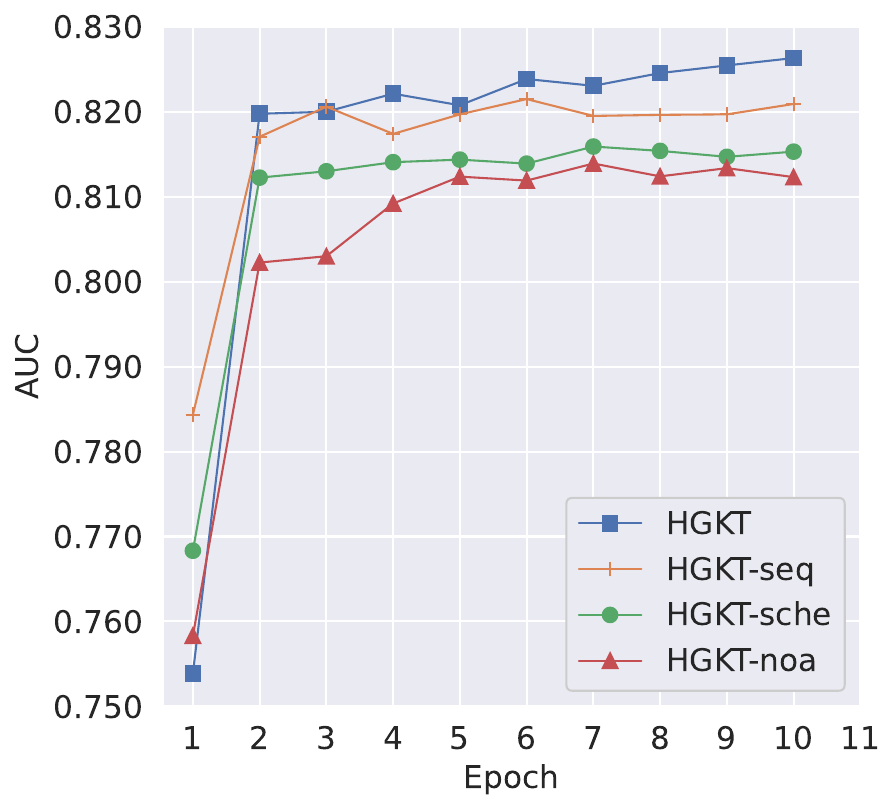}}}%
	\qquad
	\subfloat[AUC between different Attention Window 
	Size]{{\includegraphics[width=3.5cm]{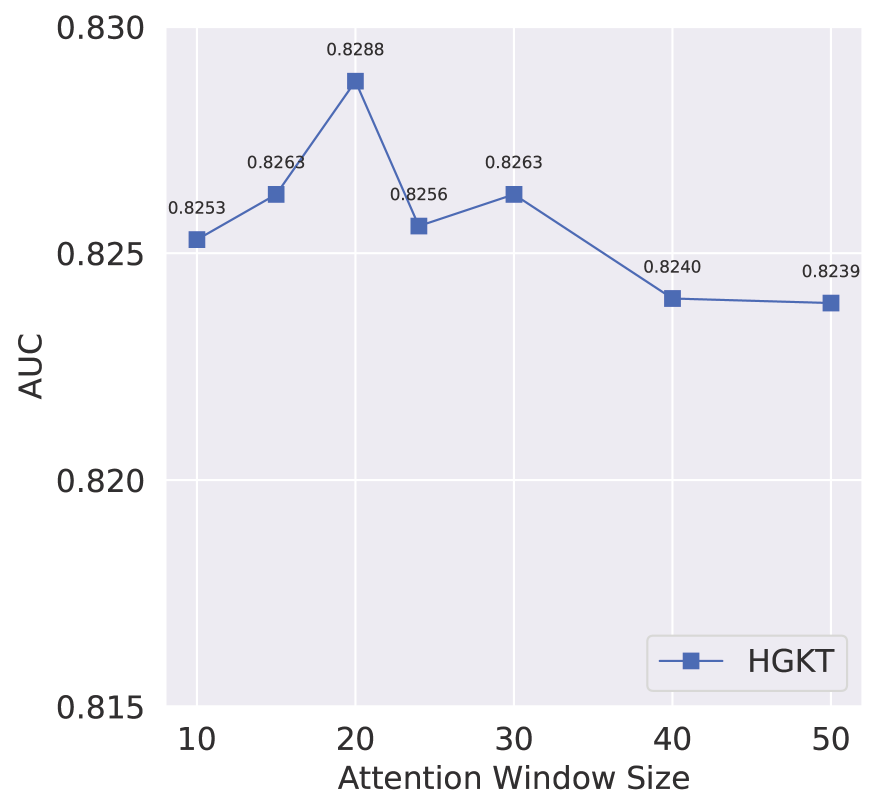}}}%
	\caption{The effectiveness of attention}% 
	\label{fig:attention_auc}%
\end{figure}

\subsubsection{\textbf{Effectiveness of Attention}} As we have clarified in Attention Mechanism section, the HGKT with attention could enhance the memory ability of the networks. The model could observe following information to predict the performance of a learner for the exercise: the performance of similar exercises in the past, learner's current knowledge state, learner's current focus. To confirm the effectiveness of those two attention mechanisms, we set up the following two experiments. Firstly, we compare the HGKT with the following three models: HGKT-noa, HGKT-sche and HGKT-seq. HGKT-noa is a no-attention version of HGKT. Similarly, HGKT-sche and HGKT-seq are variants of HGKT which only contain schema attention and sequence attention respectively. Fig.~\ref{fig:attention_auc}(a) shows the comparison results of them. From the figure, we could infer that HGKT outperforms other comparison methods. Besides, both HGKT-sche and HGKT-seq perform better than HGKT-noa, which proves that both attention mechanisms could contribute to the HGKT model. Moreover, we also conduct experiments to see the impacts of different window size of sequence attention. In the experiment, we set length of window sizes from 10 to 50 to see which size could better preserve the memory useful for KT. As Fig.~\ref{fig:attention_auc}(b) shows, the best parameter for window size is 20, which reveals that the results of nearly 20 exercises can best reflect their current learning status. This value provides a great reference for setting the number of exercises in learners' proficiency test.

%application
\begin{figure}[!ht]
	\centering
	\includegraphics[width=8cm]{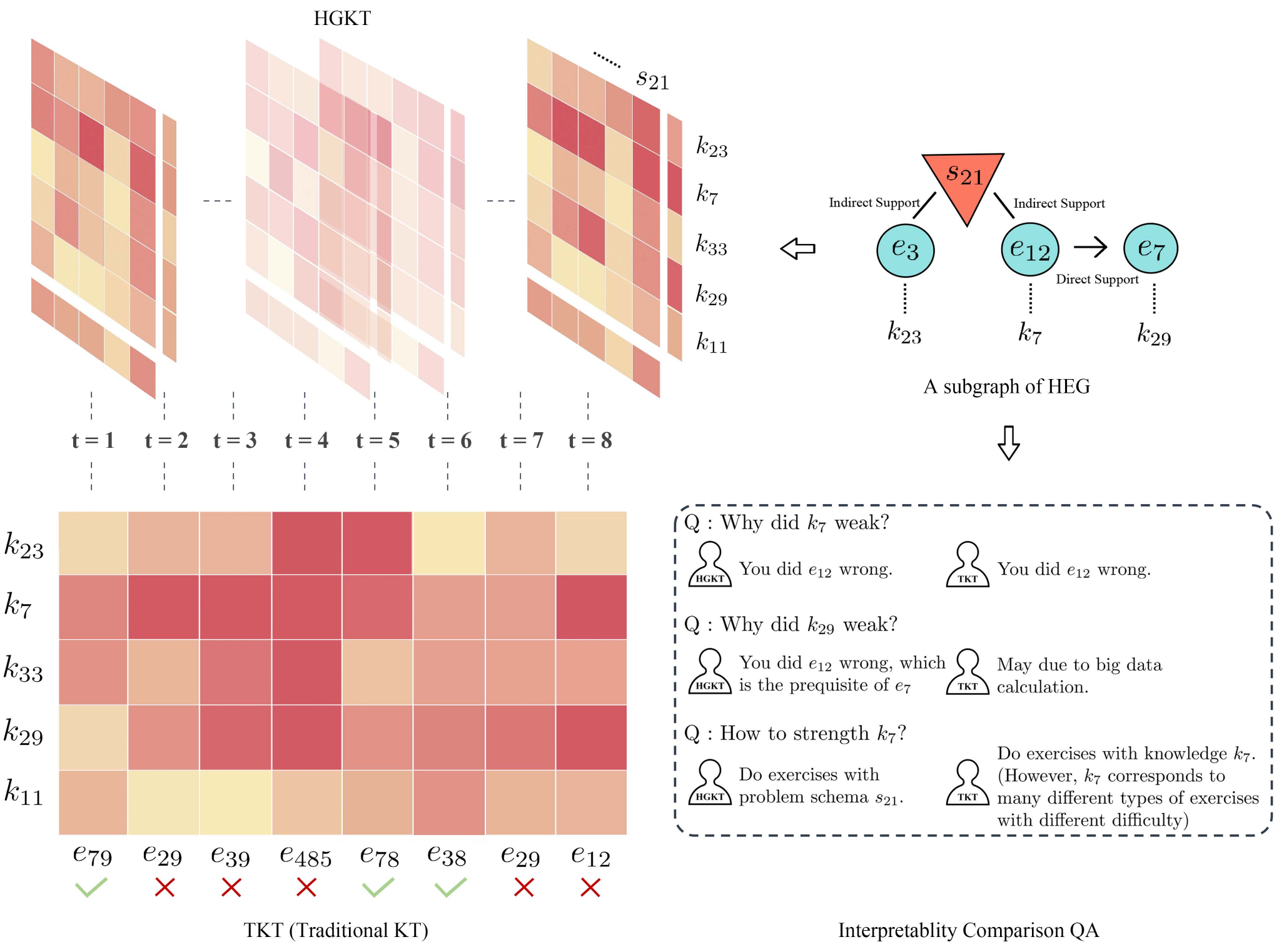}
	\caption{Each row of the diagnosis matrix corresponds to a related knowledge and each column corresponds to a related problem schema.}
	\label{fig:application}
\end{figure}

\iffalse
\renewcommand{\arraystretch}{1.2}
\begin{table*}[]
	
	\caption{P-value Comparison}
	\centering
	\resizebox{\textwidth}{!}{%
		\begin{tabular}{p{2.5cm}|llllllllll}
			\hline
			Binary Threshold & 0.1       & 0.2       & 0.3      & 
			0.4      & 
			0.5      & 0.6      & 0.7      & 0.8       & 0.9       \\
			\hline
			With-PS method   & 1.18E-164 & 5.62E-107 & 4.63E-53 & 
			3.45E-15 & 
			0.97     & 2.46E-16 & 1.44E-61 & 3.63E-141 & 8.94E-273 \\
			
			No-PS method     & 7.47E-118 & 1.91E-104 & 1.22E-87 & 
			1.53E-67 & 
			2.94E-40 & 3.92E-20 & 4.60E-03 & 1.35E-12  & 8.51E-126 \\
			\hline
		\end{tabular}%
	}
	\label{p_compare}
	
\end{table*}
\fi

\iffalse
\begin{figure}[htb]
	\centering
	\includegraphics[width=0.8\linewidth]{figures/scatter_all.eps}
	\caption{The distribution comparison between With-PS method and No-PS method}
	\label{fig:dis_compare}
\end{figure}
\fi

\section{Case Study}
The introduction of problem schema and HEG can make diagnosis more accurate and improve the interpretability of diagnosis results. Moreover, the diagnosis results generated by HGKT could be easily applied to the following two applications: (1) Learner Diagnosis Report~\cite{ragnemalm1995student,wang2019interpretable} (2) Adaptive Question Recommendation~\cite{chen2018recommendation,liu2019exploiting}. To make deep analysis about above statements, we visualize the predicted knowledge mastery of a learner during the exercising process in Fig.~\ref{fig:application}. 

From the figure we could more intuitively understand the advantages of the HGKT compared to traditional knowledge tracing diagnosis (TKT). When $t=\emph{8}$, the learner did $e_{12}$ wrong, the TKT diagnosis result shows that the color of the knowledge related to the $k_{7}$ (e.g., $k_{29}$) became darker. However, the reasons for the poor mastery of related knowledge remain unknown. In HGKT, it can be found that the transition of learners' knowledge mastery is strongly correlated with the local HEG diagram in Fig.~\ref{fig:application}(b). Specifically, the learner's mistakes in $e_{12}$ led to the poor mastery of the knowledge $k_{7}$, $k_{23}$ and $k_{29}$, which demonstrates that incorporating the direct and indirect support graph do affect the results of the diagnosis. Moreover, the HEG information could also explain the change of knowledge mastery in a way, so the learner diagnosis report could be more reliable. The QA part in the figure shows the improvement of HGKT's interpretability by comparing the explanations of HGKT and TKT for several questions. By K\&S diagnosis matrix method, learners can not only know the underlying reasons for not mastering certain knowledge (e.g., $k_{1}$, $k_{1}$), but also enhance the level of mastery of the knowledge through more targeted and accurate training advice, which reveals that HGKT framework can better be applied to adaptive question recommendation scenario.

\iffalse
\begin{table}[!ht]
	\centering
	\resizebox{0.45\textwidth}{!}{%
		\begin{tabular}{|l|l|}
			\hline
			\multirow{11}{*}{\begin{tabular}[c]{@{}l@{}}14 Proportion\\ 18 
					Addition and Subtraction Positive Decimals\\ 22 Addition 
					Whole 
					Numbers\\ 20 Addition and Subtraction Integers\\ 24 
					Addition and 
					Subtraction Fractions\\ 26 Equation Solving Two or Fewer 
					Steps\\ 31 
					Circumference\\ 44 Square Root\\ 49 Complementary and 
					Supplementary 
					Angles\\ 110 Quadratic Formula to Solve Quadratic 
					Equation\end{tabular}} & 
			\multirow{11}{*}{\begin{tabular}[c]{@{}l@{}}3 Probability of Two   
					\\ 5 Median\\ 7 Mode\\ 8 Mean\\ 9 Range\\ 10 Venn Diagram\\ 
					12 
					Circle Graph\end{tabular}} \\
			&  \\
			&  \\
			&  \\
			&  \\
			&  \\
			&  \\
			&  \\
			&  \\
			&  \\
			&  \\ \hline
		\end{tabular}%
	}
\end{table}
\fi

\section{Related Work}\label{2}

\subsubsection{Graph Neural Networks} Graph neural networks (GNN) that can model graph structured data such as social network data~\cite{hamilton2017inductive,kipf2016semi,velivckovic2017graph} or knowledge graph data~\cite{seyler2017knowledge,fan2014transition,zhang2019long} has attracted great attention recently. GNN learns node representation by transforming, propagating and aggregating node features and has been proved effective for most graph-structured data~\cite{gilmer2017neural,hamilton2017inductive}. However, the ability of current GNN architectures is limited since their graphs are inherently flat as they only propagate information across the edges of the graph and could not infer and aggregate the information in a hierarchical way~\cite{scarselli2008graph,wu2020comprehensive}. DiffPool~\cite{ying2018hierarchical} is a differentiable graph pooling method, which is first proposed to learn interpretable hierarchical representations of graph. However, it needs to train a separate cluster assignment matrix which is computation expensive. Thus, we present a variant of DiffPool to learn node representations in our paper.

\iffalse
\subsection{Text Clustering}
The purpose of text clustering is to group a set of objects in a manner where objects in the same cluster are more similar. NLP-based method usually involves three aspects~\cite{karypis2000comparison,zhao2004empirical,zhao2002evaluation,zamir1997fast}. Firstly, a suitable distance measure need to be applied to identify the proximity of two feature vectors. Secondly, text clustering requires a criterion function to get the best possible clusters and stop further processing. Thirdly, we need to propose an algorithm to optimize the criterion function. Here, the algorithms proposed could be divided into a wide variety of different types such as agglomerative clustering algorithms, partitioning algorithms, and standard parametric modeling based methods~\cite{aggarwal2012survey}. There are many commonly used methods such as K-means, Hierarchical Clustering and DBSCAN~\cite{macqueen1967some,johnson1967hierarchical,ester1996density}.

\fi

\section{Conclusion}
In this article, we prove the importance of hierarchical relations between exercises for KT tasks. To make full use of text information in exercises,  we propose a novel knowledge tracing framework HGKT which leverages the advantages of hierarchical exercise graph and sequence model with attention to enhance the ability of knowledge tracing. Besides, we come up with the concept of a K\&S diagnosis matrix that could trace the mastery of both knowledge and problem schema, which has been proved more effective and useful in industrial applications than the traditional knowledge tracing methods. Moreover, we construct a large-scale knowledge tracing dataset containing exercise text information and conduct extensive experiments to show the effectiveness and interpretability of our proposed models.

%%
%% The acknowledgments section is defined using the "acks" environment
%% (and NOT an unnumbered section). This ensures the proper
%% identification of the section in the article metadata, and the
%% consistent spelling of the heading.

\iffalse
\begin{acks}
To Robert, for the bagels and explaining CMYK and color spaces.
\end{acks}
\fi

%%
%% The next two lines define the bibliography style to be used, and
%% the bibliography file.
\bibliographystyle{ACM-Reference-Format}
\bibliography{knowledge_tracing_cikm}

%%
%% If your work has an appendix, this is the place to put it.
\appendix

\end{document}